
\magnification=1200
\tolerance=10000
\hsize 14.5truecm
\hoffset 1.25truecm
\font\cub=cmbx12
\baselineskip=24truept
\parindent=1.truecm
\def\ref{\par\noindent\hangindent 20pt}

\null
\smallskip
\hfill{DFPD 93/A/67}
\bigskip
\centerline{{\cub ~General ~Relativistic ~Dynamics ~of ~Irrotational ~Dust:}}
\medskip
\centerline{{\cub Cosmological Implications}}
\medskip
\bigskip
\noindent
\centerline{{\bf Sabino Matarrese}$^\diamond$,
{}~{\bf Ornella Pantano}$^\diamond$
{}~and ~{\bf Diego Saez}$^\star$}
\bigskip
\bigskip
\baselineskip=16truept
\noindent
$^\diamond$
Dipartimento di Fisica {\it Galileo Galilei}, Universit\`a
di Padova,

\noindent
{}~~~via Marzolo 8, I--35131 Padova, Italy

\medskip
\noindent
$^\star$
Departamento de Fisica Teorica, Universidad de Valencia,

\noindent
{}~~~Burjassot, Valencia, Spain
\vskip2.truecm
\noindent
{\bf Abstract}
The non--linear dynamics of cosmological perturbations of an irrotational
collisionless fluid is analyzed within General Relativity.
Relativistic and Newtonian solutions are compared, stressing the
different role of boundary conditions in the two theories.
Cosmological implications of relativistic effects, already
present at second order in perturbation theory, are studied and
the dynamical role of the magnetic part of the Weyl tensor is elucidated.
\bigskip
\centerline{\it{Submitted to Physical Review Letters,
October 1993}}
\bigskip
\vfill
\eject

\baselineskip=24truept

In a recent paper [1] we have shown that the General Relativistic (GR)
dynamics of a self--gravitating perfect fluid is greatly simplified under three
assumptions: {\it i}) the fluid is collisionless
(i.e. with zero pressure, $p$), {\it ii}) it has zero initial
vorticity, $\omega_{ab}$ [2] and {\it iii})
the so--called ``magnetic" part of the Weyl tensor, ${\rm H}_{ab}$,
is zero. The former two conditions are
wide enough to allow for many cosmological cases,
such as the evolution of dark matter adiabatic perturbations
generated during inflation.
The third assumption is more problematic. In
linear theory ${\rm H}_{ab}$ only contains
vector and tensor modes (e.g. Ref.[3]): if the vorticity
vanishes no vector modes are present and ${\rm H}_{ab}$ only
contains gravitational waves. Beyond linear theory the meaning
of ${\rm H}_{ab}$ is less straightforward. It is reasonable to assume that
${\rm H}_{ab}=0$ forbids at least the occurrence of gravitational waves.
This is particularly
clear in the present context, where, thanks to the absence of pressure
gradients, the motion is geodesic and, if ${\rm H}_{ab}$
also vanishes, no spatial
gradients appear in the evolution equations (apart from those contained in
convective time derivatives, which can be dropped by going to a comoving
frame): it is hard to think of any actual wave propagation with no spatial
derivatives appearing in the fluid and gravitational
evolution equations.

Following Ellis [4] we describe the dynamics directly
in terms of observable fluid and geometric quantities: the mass density
$\varrho$, the expansion scalar $\Theta$ and three traceless, flow--orthogonal
and symmetric tensors, the shear, $\sigma^\alpha_{~\beta}$,
the so--called ``electric" part of the Weyl tensor, ${\rm E}^\alpha_{~\beta}$,
describing tidal interactions of the fluid element with the
surrounding matter, and its magnetic part ${\rm H}^\alpha_{~\beta}$.
As noted in Ref.[1], if the magnetic component is switched off, all the
equations for the GR dynamics take a strictly local form:
each element evolves independently of the others. Only at the initial time
Cauchy data must be consistently given on a spatial hypersurface.
The subsequent evolution can be entirely followed in Lagrangian
form until caustic formation, when the one--to--one mapping between
fluid elements and space points is lost.
We call such a system a {\it silent universe}, in that
no information can be exchanged among different fluid elements: this is due to
the causal nature of GR, where signal exchange can only occur {\it dynamically}
via gravitational radiation and, in the case of fluids with non--zero
pressure, also via sound waves, but none of these wave modes is allowed
when $p={\rm H}_{ab}=0$.
Because of the advantages of a purely local treatment, this method [1]
has recently attracted some attention. In particular, Croudace {\it et al.}
[5] have shown the connection of the GR pancake solution [1]
with the Szekeres metric [6]; Bertschinger and Jain [7] have performed
a detailed study of the Lagrangian dynamics of fluid elements.

However, the condition ${\rm H}_{ab}=0$ cannot be taken as an exact
constraint for the general cosmological case.
It has been shown [8] that the only
solutions of Einstein
equations, with $p=\omega_{ab}={\rm H}_{ab}=0$ are either of Petrov type I, or
conformally flat, or homogeneous and anisotropic of Bianchi type I,
or {\it locally} axisymmetric (i.e. with two degenerate shear
eigenvalues)
and described by a Szekeres line--element [6].
All of these cases require some restrictions on the initial data:
the exact conditions above are not suitable to study
cosmological structure formation.
However, requiring $p=\omega_{ab}=0$ and ${\rm H}_{ab} \approx 0$
appears more feasible.
A small ${\rm H}_{ab}$ is in fact compatible with arbitrary departures from
local axisymmetry of fluid elements. This is shown by
the behaviour of perturbations around Robertson--Walker (RW):
whatever initially scalar perturbations are given,
${\rm H}_{ab}$ vanishes at first order, but not beyond. A small
value of ${\rm H}_{ab}$ allows arbitrary ratios among the shear
eigenvalues, provided the initial perturbations are small.
For general initial shapes of the
fluid elements the system will radiate gravitationally during non--linear
evolution. However, fully GR numerical computations [9] have shown that
only a negligible fraction (less than $1 \%$) of the total energy is carried
away in the form of gravitational radiation, during the non--linear collapse
of collisionless ellipsoids.
In spite of these facts, as our calculations below demonstrate,
a non--zero ${\rm H}_{ab}$ allows for the influence
of the surrounding matter on the
evolution of fluid elements. Although this signal travels at finite
speed, for perturbations on scales much smaller than the horizon it
effectively appears as an instantaneous Newtonian feature.
One might wonder whether during the late
phases of collapse, when local axisymmetry is expected to be established, the
environmental influence on the evolving fluid element can be
neglected and the ${\rm H}_{ab} = 0$ condition restored.

\medskip
\noindent
{\it General relativistic dynamics} --
To describe our system we start from the equations of Ref.[4]. We always work
in the comoving synchronous
gauge $ds^2= - dt^2+a^2(t) {\tilde h}_{\alpha\beta} dq^\alpha dq^\beta$,
where $a = A t^{2/3}$, as for a flat, matter--dominated RW
model (our ``background" solution).
For computational convenience we introduce suitably rescaled quantities: a
scaled density fluctuation $\Delta \equiv (6\pi G t^2 \varrho -1)/a$,
a peculiar expansion scalar $\vartheta = (3t/2a)(\Theta - 2/t)$, a
traceless shear tensor $s^\alpha_{~\beta} \equiv (3t/2a)
\sigma^\alpha_{~\beta}$ and a traceless tidal tensor
$e^\alpha_{~\beta} \equiv (3t^2/2a) {\rm E}^\alpha_{~\beta}$.
These quantities can be grouped in two space--like tensors:
the {\it velocity gradient} tensor $\vartheta^\alpha_{~\beta} \equiv
s^\alpha_{~\beta} + {1\over 3}
\delta^\alpha_{~\beta} \vartheta$, related to the
covariant derivatives of the peculiar velocity field;
the {\it peculiar gravitational field} tensor
$\Delta^\alpha_{~\beta} \equiv e^\alpha_{~\beta} + {1 \over 3}
\Delta \delta^\alpha_{~\beta}$. We also scale the magnetic tensor
as ${\cal H}^\alpha_{~\beta} \equiv (3t^2/2a) {\rm H}^\alpha_{~\beta}$.
The dynamical equations for the fluid and the gravitational field are
$$
\eqalignno{
{\dot \vartheta}^\alpha_{~\beta} = &
-{3 \over 2 a} (\vartheta^\alpha_{~\beta}
+ \Delta^\alpha_{~\beta}) - \vartheta^\alpha_{~\gamma}
\vartheta^\gamma_{~\beta} \ , &
(1) \cr
{\dot \Delta}^\alpha_{~\beta} = &
- { 1 \over a} (\vartheta^\alpha_{~\beta}
+ \Delta^\alpha_{~\beta})
- 2(\vartheta \Delta^\alpha_{~\beta} + \Delta \vartheta^\alpha_{~\beta}) +
{5 \over 2} \Delta^\alpha_{~\gamma} \vartheta^\gamma_{~\beta}
+ {1 \over 2} \Delta^\gamma_{~\beta}\vartheta^\alpha_{~\gamma} + & \cr
& + \delta^\alpha_{~\beta}(\Delta \vartheta -
\Delta^\gamma_{~\delta}\vartheta^\delta_{~\gamma}) + {3 t \over 4 a^2}
\tilde h_{\beta\eta} \bigl( \tilde \eta^{\eta\gamma\delta}
{\cal H}^\alpha_{~\gamma};_{\delta} + \tilde \eta^{\alpha\gamma\delta}
{\cal H}^\eta_{~\gamma};_{\delta} \bigr)\ , &
(2) \cr
{\dot {\cal H}}^\alpha_{~\beta}
= & - {1 \over a} {\cal H}^\alpha_{~\beta}
- 2 \vartheta {\cal H}^\alpha_{~\beta} - \delta^\alpha_{~\beta}
\vartheta^\gamma_{~\delta} {\cal H}^\delta_{~\gamma} +
{5 \over 2} {\cal H}^\alpha_{~\gamma} \vartheta^\gamma_{~\beta}
+ {1 \over 2} {\cal H}^\gamma_{~\beta}\vartheta^\alpha_{~\gamma} - & \cr
& - {3 t \over 4 a^2} \tilde h_{\beta\eta}
\bigl( \tilde \eta^{\eta\gamma\delta}
\Delta^\alpha_{~\gamma};_{\delta} + \tilde \eta^{\alpha\gamma\delta}
\Delta^\eta_{~\gamma};_{\delta} \bigr)\ , &
(3) \cr}
$$
where the dot denotes partial differentiation with respect to the scale
factor $a$ and $\tilde \eta^{\alpha\gamma\delta}$ is the
Levi--Civita tensor relative to the metric $\tilde h_{\alpha\beta}$:
$\tilde \eta^{\alpha\beta\gamma} = \tilde h^{-1/2}
\varepsilon^{\alpha\beta\gamma}$, with $\varepsilon^{123}=1$.
The metric tensor evolves according to
${1 \over 2} \tilde h^{\alpha\gamma} {\dot {\tilde h}}_{\gamma\beta} =
\vartheta^\alpha_{~\beta}$.

The above tensors have to satisfy the constraints [4]
$$
\eqalignno{
&  {\vartheta_\alpha^{~\beta}};_\beta = \vartheta,_\alpha \ , &
(4) \cr
&  {\Delta_\alpha^{~\beta}};_\beta = \Delta,_\alpha - {2 a^2 \over 3 t}
\tilde h_{\alpha\mu} \tilde h_{\beta\nu} \tilde \eta^{\mu\lambda\gamma}
\vartheta^\nu_{~\lambda} {\cal H}^\beta_{~\gamma} \ , &
(5) \cr
& {\cal H}_\alpha^{~\beta};_\beta = {2 a^2 \over 3 t} \tilde h_{\alpha\mu}
\tilde h_{\beta\nu} \tilde \eta^{\mu\lambda\gamma}
\vartheta^\nu_{~\lambda} \Delta^\beta_{~\gamma} \ , &
(6) \cr
& {\cal H}^\alpha_{~\beta} = { t \over 2 a} \tilde h_{\beta\mu} \bigl(
\tilde \eta^{\mu\gamma\delta} \vartheta_\gamma^{~\alpha};_\delta +
\tilde \eta^{\alpha\gamma\delta} \vartheta_\gamma^{~\mu};_\delta \bigr)
\ . &
(7) \cr}
$$
All these are fulfilled at the linear level [3] by
growing--mode scalar initial conditions [1]: $\Delta^\alpha_{~\beta} (a_0) =
- \vartheta^\alpha_{~\beta}(a_0) = \varphi_0,^\alpha_{~\beta}$,
where the scalar $\varphi_0$, an arbitrary function of the space
coordinates $q^\alpha$, is the initial peculiar gravitational potential,
related to Bardeen's gauge--invariant $\Phi_H$ [10]
by $\varphi_0 = - (3 / 2 A^3) \Phi_H$.
These initial conditions correspond to the ``seed"
metric $\tilde h_{\alpha\beta} =
\delta_{\alpha\beta}(1 - {20 \over 9}
A^3 \varphi_0) - 2 a \varphi_0,_{\alpha\beta}$,
and imply vanishing initial ${\cal H}^\alpha_{~\beta}$
(the constant mode, $\propto A^3 \varphi_0 \ll 1$, can be neglected in
practice, compared to the growing mode
$\propto a\varphi_0,_{\alpha\beta}$).

The Lagrangian dynamics is determined by Eqs.(1), (2) and (3) plus
the initial data. One obtains a local Eulerian description
of the fluid [1], using the ``generalized Hubble law" [4].
We have ${\dot \xi}^\alpha = \vartheta^\alpha_{~\beta}\xi^\beta$,
where $a \xi^\alpha$ is the
infinitesimal spatial displacement of neighbouring elements.
The matrix connecting the Eulerian coordinates $x^\alpha$ with the
Lagrangian ones $q^\beta$ is the Jacobian
$J^\alpha_{~\beta} \equiv \partial x^\alpha / \partial q^\beta \equiv
\delta^\alpha_{~\beta} + {\cal D}^\alpha_{~\beta}$,
where ${\cal D}^\alpha_{~\beta}$ is the (symmetric) deformation tensor.
Taking $\xi^\alpha = d x^\alpha = J^\alpha_{~\beta}
\xi_{(0)}^{~\beta}$, where $\xi_{(0)}^{~\beta} = dq^\beta$ represent the
initial (i.e. Lagrangian) infinitesimal displacements,
one gets
${\dot {\cal D}}^\alpha_{~\beta}  = \vartheta^\alpha_{~\beta} +
\vartheta^\alpha_{~\gamma}{\cal D}^\gamma_{~\beta}$,
formally solved by ${\cal D}^\alpha_{~\beta}(a) = \exp\int_{a_0}^a
d \overline a \vartheta^\alpha_{~\beta}(\overline a )
- \delta^\alpha_{~\beta}$.
Once the Jacobian is known one gets the metric
as $\tilde h_{\alpha\beta} = \tilde h_{\gamma\delta}(a_0) J^\gamma_{~\alpha}
J^\delta_{~\beta}$.
As shown in Refs.[8,1], if ${\cal H}^\alpha_{~\beta}=0$,
the tensors $\vartheta^\alpha_{~\beta}$, $\Delta^\alpha_{~\beta}$,
$\tilde h_{\alpha\beta}$ commute and they can be diagonalized
simultaneously. In such a case, Eqs.(1) and (2) can be reduced
to six first order equations for the six
eigenvalues of $\vartheta^\alpha_{~\beta}$ and $\Delta^\alpha_{~\beta}$.
Along the local principal axes we can set ${\tilde h}_{\alpha\beta} =
\delta_{\alpha\beta}
{\tilde h}_\beta$ and $\vartheta^\alpha_{~\beta} = \delta^\alpha_{~\beta}
\vartheta_\beta$ and get ${\tilde h}_\alpha(a) = {\tilde h}_\alpha(a_0)
\exp 2 \int_{a_0}^a d \overline a \vartheta_\alpha( \overline a )$.
In the locally axisymmetric case, i.e. when two eigenvalues of
$\varphi_0,^\alpha_{~\beta}$ coincide, a relation exists with particular
Szekeres solutions [6].

\medskip
\noindent
{\it Newtonian dynamics} --
The equations which govern the non--linear dynamics of a collisionless fluid
in Newtonian Theory (NT) for an expanding universe [11] can be
written in suitably rescaled form as (e.g. Ref.[12])
$$
\eqalignno{
& {\dot u}^\alpha + u^\beta u^\alpha,_\beta = - {3 \over 2 a} (u^\alpha +
\varphi,^\alpha) \ , &
(8) \cr
& {\dot \Delta} +
u^\beta \Delta,_\beta = - {1 \over a} (u^\beta,_\beta + \Delta) -
\Delta u^\beta,_\beta \ , &
(9) \cr
& \varphi,^\beta_{~\beta} = \Delta \ , &
(10) \cr}
$$
where $\varphi$ is the peculiar gravitational potential. Differentiating the
Euler equation (8), defining the symmetric tensors
$\vartheta^\alpha_{~\beta} \equiv u^\alpha,_\beta$, with $u^\alpha=dx^\alpha /
da$, and $\Delta^\alpha_{~\beta} \equiv \varphi,^\alpha_{~\beta}$, and
adopting a Lagrangian description, one recovers Eq.(1),
while the continuity equation (9) coincides with the trace of Eq.(2).
It is clear that the NT is degenerate,
as it provides only one equation to determine the tensor
$\Delta^\alpha_{~\beta}$: any traceless tensor added to the r.h.s. of Eq.(2),
leaves the NT equations unchanged. In order to
completely determine the evolution of the gravitational field tensor
$\Delta^\alpha_{~\beta}$ one has to resort to its definition
in terms of the potential $\varphi$, i.e. to a {\it non--local}
theory. Because of the intrinsic non--locality of NT (the Poisson equation
(10) is an elliptic, constraint equation) one needs boundary conditions to
determine the dynamics: contrary to the GR equations,
initial data are not enough.
It is well known (e.g. Ref.[4]) that the
lack of evolution equations for the traceless part of the gravitational
tensor, $e^\alpha_{~\beta}$, implies that the
NT adds spurious solutions which would be
discarded by the full GR system.

\medskip
\noindent
{\it Beyond the Zel'dovich approximation} --
In order to see the behaviour of the GR solutions
and evaluate the role of the magnetic term we construct
a second order Lagrangian perturbation expansion
in the amplitude of the fluctuations around RW. It will prove useful to define
the two quantities $\mu_1 \equiv \varphi_0,^\gamma_{~\gamma} = \lambda_1 +
\lambda_2 + \lambda_3$ and $\mu_2 = {1 \over 2}(\varphi_0,^\gamma_{~\gamma}
\varphi_0,^\delta_{~\delta} - \varphi_0,^\gamma_{~\delta}
\varphi_0,^\delta_{~\gamma}) =
\lambda_1\lambda_2 + \lambda_1\lambda_3 + \lambda_2\lambda_3$,
where $\lambda_\alpha$ are the local eigenvalues of the symmetric tensor
$\varphi_0,^\alpha_{~\beta}$. One immediately obtains the
traces, $\vartheta = - \mu_1 + a(- \mu_1^2 + {8 \over 7} \mu_2)$ and
$\Delta = \mu_1 + a(\mu_1^2 - {4 \over 7} \mu_2)$, which
coincide with those obtained in Lagrangian second order NT [13].
After very lengthy calculations we obtain
$$
\vartheta^\alpha_{~\beta} \equiv - \varphi_0,^\alpha_{~\beta} + {a \over 7}
\bigl( - 12 \mu_1 \varphi_0,^\alpha_{~\beta} +
6 \mu_2 \delta^\alpha_{~\beta} + 5
\varphi_0,^\alpha_{~\gamma} \varphi_0,^\gamma_{~\beta} \bigr) +
\chi^\alpha_{~\beta} \ ,
\eqno(11)
$$
(here indices are raised by the Kronecker symbol), having kept only growing
modes. The expressions for $\Delta^\alpha_{~\beta}$ and
${\cal H}^\alpha_{~\beta}$ will not be reported here for
shortness. The traceless tensor $\chi^\alpha_{~\beta}$, representing the
contribution due to the magnetic part, has zero divergence:
$\chi_\beta^{~\alpha},_\alpha=0$. It can be written
as a convolution $\chi^\alpha_{~\beta}({\bf q},a) = \int d^3 q'
S^\alpha_{~\beta} ({\bf q'})
f(\vert {\bf q} - {\bf q'}\vert,a)$, of the source
${\cal S}^\alpha_{~\beta} =
{\mu_2},^\alpha_{~\beta} - \nabla^2 (2 \mu_1
\varphi_0,^\alpha_{~\beta} - 2 \varphi_0,^\alpha_{~\gamma}
\varphi_0,^\gamma_{~\beta} - \delta^\alpha_{~\beta} \mu_2 )$
with the function $f$, whose Fourier transform $\hat f(k)$ satisfies
the equation,
$$
\hat f''' + {9 \over \tau} \hat f'' +
{12 \over \tau^2} \hat f' + k^2
\biggl( \hat f' + {3 \over \tau} \hat f \biggr)
= {10 A^3 \tau \over 21} \ ,
\eqno(12)
$$
where a prime denotes differentiation with respect to the conformal time
$\tau = (3/A) t^{1/3}$. The initial conditions are
$\hat f(\tau_0) = \hat f'(\tau_0) = \hat f''(\tau_0) = 0$.
Asymptotic solutions of
Eq.(12), confirmed by a numerical check, are
$\hat f \approx 2 A^3 \tau^2  /21 k^2$ for $k\tau \gg 1$
and $\hat f \approx A^3 \tau^4 /378$ for $k\tau \ll 1$.
Performing a second--order
expansion for the deformation tensor and defining
${\cal D}^\alpha_{~\beta} \equiv - a \varphi_0,^\alpha_{~\beta} + a^2
\psi^\alpha_{~\beta}$, we find
$$
\psi^\alpha_{~\beta} = {3 \over 7} \biggl( - 2 \mu_1
\varphi_0,^\alpha_{~\beta} + \mu_2 \delta^\alpha_{~\beta} + 2
\varphi_0,^\alpha_{~\gamma} \varphi_0,^\gamma_{~\beta}\biggr) + {1 \over a^2}
\int_{a_0}^a d \overline a \chi^\alpha_{~\beta} \ ,
\eqno(13)
$$
with trace $\psi^\alpha_{~\alpha} = - {3 \over 7} \mu_2$.
The symmetric tensor $\psi^\alpha_{~\beta}$ provides the second order
correction to the deformation tensor, whose first order is
the kinematical Zel'dovich approximation.
The metric tensor reads
$$
\tilde h_{\alpha\beta} = \delta_{\alpha\beta}
- 2 a \varphi_0,_{\alpha\beta} +
{a^2 \over 7} \biggl(19 \varphi_0,_{\alpha\gamma} \varphi_0,^\gamma_{~\beta}
- 12 \mu_1 \varphi_0,_{\alpha\beta} +
6 \mu_2 \delta_{\alpha\beta} \biggr) +
\int_{a_0}^a d \overline a \chi_{\alpha\beta} \ .
\eqno(14)
$$

We then have
$dx^\alpha = dq^\alpha - a \varphi_0,^\alpha_{~\beta} dq^\beta +
a^2 \psi^\alpha_{~\beta} d q^\beta$.
In NT one would write
the same formal expression, but the irrotationality
condition would lead to $\psi^\alpha_{~\beta} = \psi,^\alpha_{~\beta}$,
with the potential $\psi$ satisfying the second order Poisson equation [13]
$\nabla^2 \psi = - {3 \over 7} \mu_2$, which is
consistent with the trace of the GR equation.
In other words, the NT eigenvalues $\nu_\alpha$ of $\psi^\alpha_{~\beta}$
only need to satisfy the
condition $\sum_\alpha \nu_\alpha = - {3 \over 7} \mu_2$.
In order to get the complete information on the single
$\nu_\alpha$'s one needs the NT definition of
$\psi^\alpha_{~\beta}$
as $\psi,^\alpha_{~\beta}$, i.e. a non--local information.
The GR $\nu_\alpha$'s also solve the NT equations,
but the reverse is not necessarily true: it
depends upon the boundary conditions used in solving
Poisson's equation.

\medskip
\noindent
{\it Inside the horizon} --
Suppose that the source, hence $\varphi_0,^\alpha_{~\beta}$, has some
typical scale of variation $\ell$, i.e. $\ell \sim \varphi_0,^\alpha_{~\beta}
/\varphi_0,^\alpha_{~\beta\gamma}$. If $\ell \ll \tau$ we find
$\vartheta^\alpha_{~\beta} \approx - \varphi_0,^\alpha_{~\beta} -
a \varphi_0,^\alpha_{~\gamma} \varphi_0,^\gamma_{~\beta} +
2 a \psi,^\alpha_{~\beta}$.
The second order deformation tensor reduces to $\psi^\alpha_{~\beta}=
\psi,^\alpha_{~\beta}$, while the metric reads
$\tilde h_{\alpha\beta} = \delta_{\alpha\beta}
- 2 a \varphi_0,_{\alpha\beta} + a^2 \psi,_{\alpha\beta}$.
All these expressions coincide with those of second
order NT and can be obtained from the $c \to \infty$ limit
of Eq.(12). The scalar $\psi$ carries information on the
influence of the surrounding matter on the dynamics of fluid elements.
Note that $\psi,^\alpha_{~\beta}$ produces a tilt of the principal
axes of the first--order deformation tensor, $\varphi_0,^\alpha_{~\beta}$.

\medskip
\noindent
{\it Outside the horizon} --
When $\ell \gg \tau$, $\chi^\alpha_{~\beta} \approx
(3 t^2 / 14 a) {\cal S}^\alpha_{~\beta}$, and the  contribution to
$\vartheta^\alpha_{~\beta}$ due to the magnetic term becomes negligible.
The relevant expressions can be obtained from Eqs.(11), (13) and (14) with
$\chi^\alpha_{~\beta} \approx 0$.
Perturbations with size greater than the Hubble
radius evolve as a separate {\it silent} universe: spatial gradients play no
role in this case. However, these local GR
effects have little cosmological implications, since perturbations on
super--horizon scales usually have very small amplitude, and
a linear approximation is sufficient. Nevertheless, there are a number
of formal consequences, which is worth mentioning.
One of these is the {\it absence of 2D solutions}.
If one eigenvalue of $\varphi_0,^\alpha_{~\beta}$,
e.g. $\lambda_3$, vanishes everywhere, the NT, with suitable boundary
conditions, implies
$\vartheta_3(a)=0$ or $x_3(a)=q_3$, i.e. no motion along the third
axis.
This is referred as ``two--dimensional" (2D) gravitational clustering.
As far as the second order deformation tensor is concerned, one would have
$\nu_1+\nu_2 = - {3 \over 7} \mu_2$, with $\mu_2=\lambda_1\lambda_2$, and
$\nu_3=0$.
In the GR case, instead, we find $\nu_1 = \nu_2 = - \nu_3 = - {3 \over 7}
\mu_2$, and $\vartheta_3(a) \neq 0$ for $a \neq a_0$.
The motion dynamically impressed along the third axis soon becomes of the same
order of magnitude as that in the other directions. This effect
is due to the tide--shear coupling term
$\delta^\alpha_{~\beta} (\vartheta \Delta - \Delta^\gamma_{~\delta}
\vartheta^\delta_{~\gamma})$ in the evolution equation for
$\Delta^\alpha_{~\beta}$, which reduces to $- 2 \mu_2 \delta^\alpha_{~\beta}$
to lowest order.
The only case when this coupling disappears is when
two $\lambda_\alpha$'s simultaneously vanish, i.e. for planar symmetry.
Therefore $\vartheta_3(a)=0$ is not an exact
solution of the GR equations, unless another $\vartheta_\alpha$ also
vanishes. As an example,
no axisymmetric configurations without motion along the symmetry axis are
allowed.

This discussion leads to the main issue: the general non--linear dynamics
of fluid elements.
So far, two analytical solutions of our system are known: for
planar configurations, $\lambda_1=\lambda_2=0$, one
recovers the Zel'dovich pancake solution, as shown in
Ref.[1]; for exactly spherical configurations,
$\lambda_1=\lambda_2=\lambda_3$, the local solution is the well--known top--hat
model (e.g. Ref.[11]). Croudace {\it et al.} [5] looked for
solutions representing attractors
among the trajectories of our system with zero magnetic tensor. They found
that both spherical collapse and a
perfect pancake are repellers for general initial conditions, and
argued that the pancake instability is probably due to having
disregarded the contribution of ${\cal H}^\alpha_{~\beta}$. On the other hand,
Bertschinger and Jain [7] have shown that the
instability of the pancake solution is caused by the tide--shear
coupling in the evolution of the tide, which tends to destabilize
the pancake solution (for general initial conditions) but would
stabilize prolate configurations. For vanishing ${\cal H}^\alpha_{~\beta}$,
a strongly prolate
spindle with expansion along its axis is the generical outcome of collapse,
except for specific initial conditions corresponding
to exactly spherical or planar configurations. Our analysis shows
that the dynamical effect causing preferential collapse to expanding spindles
in the ${\cal H}^\alpha_{~\beta}=0$ case is the GR tide--shear coupling in the
tide evolution equation. This term is not present in NT,
although it is compatible with its equations.
This is further illustrated by the
collapse of an infinite  homogeneous ellipsoid ($\ell \to \infty$),
which is described by our equations with zero ${\cal H}^\alpha_{~\beta}$.
As well known, the
NT dynamics favours the formation of oblate spheroids (e.g. Ref.[14]),
pancake--like objects
with one collapsing axis and the other two tending to a finite size
(apart from initial conditions corresponding to
an initial prolate spheroid).
The GR collapse (e.g. Ref.[15]) favours the formation of prolate
spheroids, collapsing filaments with expansion along their symmetry axis.
However, as our second order calculations show, the evolution of
fluid elements as isolated ellipsoids does not apply to perturbations on scales
smaller than the Hubble radius: here non--local effect play a fundamental role.
The actual non--linear dynamics would generally result
from the competition of the {\it local} GR tide--shear coupling, causing
pancake instability, and the {\it non--local} environmental influence,
carried by the magnetic part of the Weyl tensor. During the early deviations
from linear evolution, as described by Lagrangian second order perturbation
theory, the latter effect dominates; however, extending this conclusion to the
late strongly non--linear phases would require further study.

Finally, as a result of this analysis, we are able to calculate how many
gravitational waves are produced within a second order approximation
(remember that at this order the magnetic tensor is
traceless and transverse, so it is related to gravitational radiation [3]).
Outside the horizon ${\rm H}_{\alpha\beta} \approx (4 a^3 / 7 t)
(\varepsilon_{\alpha\gamma\delta} \varphi_0,^\delta_{~\beta}
\varphi_0,^\gamma_{~\nu} + \varepsilon_{\beta\gamma\delta}
\varphi_0,^\delta_{~\alpha} \varphi_0,^\gamma_{~\nu}),^\nu$,
while inside the horizon ${\rm H}_{\alpha\beta}$ decays as $1/a$.
Only a tiny amount of gravitational waves is produced
on sub--horizon scales at this level, however their dynamical role is far from
being negligible!

\medskip
\baselineskip=18truept
\noindent
{\bf Acknowledgments}
M. Bruni is acknowledged for useful discussions.
This work was partially supported by Italian MURST. DS thanks the
Conselleria de Cultura, Educacio i Ciencia de la Generalitat Valenciana
and the Spanish DGICYT project PB90-0416 for support.

\bigskip
\noindent
{\bf References}
\bigskip

\ref {1. ~S. Matarrese, O. Pantano and D. Saez,
Phys. Rev. {\bf D47}, 1311 (1993).}

\ref {2. ~Latin indices label space--time coordinates, $(0,1,2,3)$,
greek indices spatial ones, $(1,2,3)$. Commas are used
for partial derivatives, semicolons for covariant ones.}

\ref {3. ~M. Bruni, P.K.S. Dunsby and G.F.R. Ellis, Astrophys. J. {\bf
395}, 34 (1992)}

\ref {4. ~G.F.R. Ellis, in {\it General Relativity and Cosmology}, ed. R.K.
Sachs (Academic Press, New York, 1971).}

\ref {5. ~K.M. Croudace, J. Parry, D.S. Salopek and J.M. Stewart,
Astrophys. J., in press (1993).}

\ref {6. ~P. Szekeres, Commun. Math. Phys. {\bf 141}, 55 (1975).}

\ref {7. E. Bertschinger and B. Jain, MIT preprint (1993); see also
E. Bertschinger, MIT preprint (1993).}

\ref {8. ~A. Barnes and R.R. Rowlingson, Class. Quantum Grav. {\bf 6},
949 (1989).}

\ref {9. ~S.L. Shapiro and S.A. Teukolsky, Phys. Rev. Lett. {\bf 66},
994 (1991).}

\ref {10. J.M. Bardeen, Phys. Rev. {\bf D22}, 1882 (1980).}

\ref {11. P.J.E. Peebles, {\it The Large Scale Structure of the
Universe} (Princeton University Press, Princeton, 1980).}

\ref {12. S. Matarrese, F. Lucchin, L. Moscardini and D. Saez,
Mon. Not. R. Astr. Soc. {\bf 259}, 437 (1992).}

\ref {13. T. Buchert, Astron. Astrophys. {\bf 223}, 9 (1989), Mon. Not. R.
Astr. Soc. {\bf 254}, 729 (1992); F. Moutarde, J.--M. Alimi, F.R. Bouchet, R.
Pellat and A. Ramani, Astrophys. J. {\bf 382}, 377 (1991);
M. Gramann, Astrophys. J. {\bf 405}, L47 (1993);
M. Lachi\`eze--Rey, Astrophys. J. {\bf 408}, 403 (1993).}

\ref {14. S.D.M. White and J. Silk, Astrophys. J. {\bf 231}, 1
(1979).}

\ref {15. Ya.B. Zel'dovich and I.D. Novikov, {\it The Structure and Evolution
of the Universe} (The University of Chicago Press, Chicago, 1983) Chapter 19.}
\vfill\eject
\bye